# Continuous-wave backward frequency doubling in periodically poled lithium niobate

S. Stivala,[1] A. C. Busacca,[1] L. Curcio,[1] R. L. Oliveri,[1] S. Riva-Sanseverino,[1] and Gaetano Assanto[2,a]

[1]*DIEET, University of Palermo, Viale delle Scienze, Bldg. no. 9, 90128 Palermo, Italy*
[2]*Nonlinear Optics and OptoElectronics Laboratory (NooEL), University "Roma Tre", Via della Vasca Navale 84, 00146 Rome, Italy*

We report on backward second-harmonic-generation in bulk periodically poled congruent lithium niobate with a 3.2 μm period. A tunable continuous-wave Ti:sapphire laser allowed us exciting two resonant quasi-phase-matching orders in the backward configuration. The resonances were also resolved by temperature tuning and interpolated with standard theory to extract relevant information on the sample.

Periodic domain inversion by electric field poling of nonlinear ferroelectric crystals is a well known technique to achieve efficient three-wave interactions via quasi-phase matching (QPM).[1,2] Backward-second-harmonic-generation (BSHG), an up-conversion process with the fundamental frequency (FF) and its second harmonic propagating in opposite directions, was introduced by Harris in Ref. 3 and studied in detail in Refs. 4 and 5. Its first experimental observations via QPM were reported in multilayer semiconductor heterostructures[6] and in periodically poled (PP) congruent lithium niobate (LN),[7,8] exploiting high orders of resonance in gratings of periods of 3–4 μm. Recent technological improvements allowed fabricating submicron-period gratings in potassium titanyl phophate (KTP) waveguides and bulk, observing BSHG at lower QPM orders.[9,10]

Among the configurations to be implemented in counter-propagating configurations, Conti et al.[11,12] proposed purely nonlinear gap-solitons and cavityless oscillators, Canalias et al.[13] demonstrated a mirrorless optical parametric oscillator in bulk KTP. BSHG by a continuous wave (cw) input in PPLN was previously reported only in Ref. 14, even though this type of excitation seems to be more appropriate for fine tuning the FF wavelength to the very sharp resonance(s) characteristic of this parametric process.

Herein we describe fabrication and cw nonlinear characterization of a 3.2 μm-period PPLN bulk sample. After discussing first-order QPM for *forward* SHG, we report our results on *backward* SHG with two different resonant orders, using thermo-optic adjustments as well as FF wavelength tuning. The experimental results are in good agreement with the predictions and, through standard modeling, permit a diagnostic assessment of the technologic shortcomings.

The sample was prepared in a polished optical-grade wafer of congruent lithium niobate 500 μm thick. By standard photolithography a one-dimensional grating (15 mm long and 10 mm wide) of period 3.2 μm was defined over a length of 3 mm on a 2-μm-thick film of positive photoresist (Shipley S1813) spin-coated on the −Z facet. After development, the pattern was soft-baked overnight at 90 °C and hard-baked for 3 h at 130 °C. The curing temperature was raised gradually in order to avoid problems deriving from the pyroelectric effect. This treatment provided a better adhesion between the photoresist and the substrate while improving insulation. For the poling, the sample was contacted with gel-electrolyte layers and subjected to high voltage pulses supplied by an amplified (Trek) waveform generator (Agilent). In order to exceed the LN coercive field and obtain a charge-controlled domain inversion, we applied single 1.3 kV pulses over a 10 kV offset for appropriate time intervals. With this approach, the inverted domains enucleated from the −Z facet in the region under the electrodes and extended toward the +Z facet before coalescence occurred.[15,16] Finally, the end-facets of the chip were polished before proceeding with the characterization.

We used a cw 40 GHz line-width Ti:sapphire laser tunable in wavelength between 860 and 1000 nm; with an f=11 mm lens the TEM$_{00}$ beam was gently focused to a waist of 30 μm in the middle of the poled area, keeping the crystal at a constant temperature of 195.0 °C to reduce or avoid the risk of photorefractive damage. The FF beam was mechanically chopped at 133 Hz and the generated SH was collected by a photomultiplier tube for synchronous detection; the residual pump was eliminated with the aid of a dichroic mirror and selective filters. First we studied the SH generated forward via first-order QPM by light at $\lambda_{FF}$=862.33 nm. We spectrally resolved the QPM-SHG resonance by varying either the wavelength [Fig. 1(a)] or the temperature [Fig. 1(b)] while keeping the other parameters constant. We verified the quadratic growth of the SH versus FF input power [Fig. 1(c)] and interpolated the experimental data with the standard expression of the SHG efficiency conversion.[2,17] Assuming a good overlap between the FF beam and the poled domains, using the value of 25.2 pm/V for the quadratic element $d_{33}$ nonlinearly coupling FF and SH electric field components along the optic axis,[18] we estimated a QPM duty-cycle around 75:25 and a period of 3.2 μm. The measured full-width at half-maximum of 0.24 nm corresponded to a QPM interaction over 3 mm, i.e., the total length of the grating with a uniform period of nominal value.

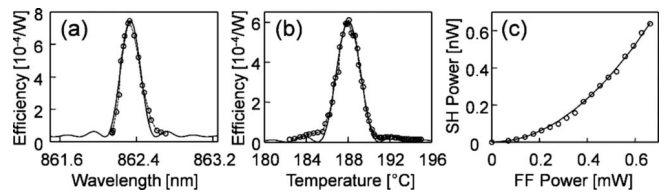

FIG. 1. Normalized SHG efficiency (a) vs FF wavelength and (b) temperature for an input power of 0.7 mW; (c) parabolic trend of the generated SH vs FF input power at the QPM resonant wavelength. The solid lines are from the model; the dashed lines connecting data points are guides to the eye S

---
[a]Electronic mail: assanto@uniroma3.it

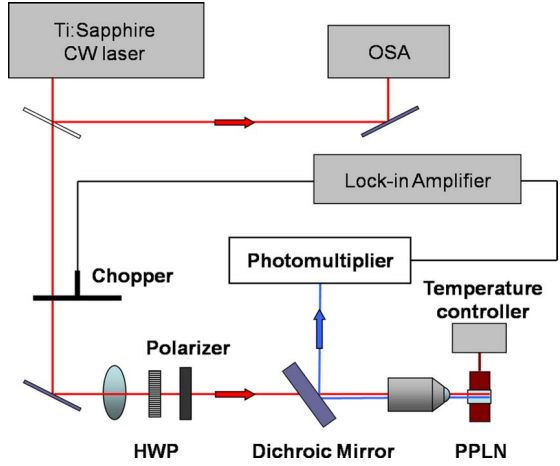

FIG. 2. (Color online) Set-up for BSHG measurements. HWP stands for half wave plate and OSA for optical spectrum analyzer. Additional filters are not shown.

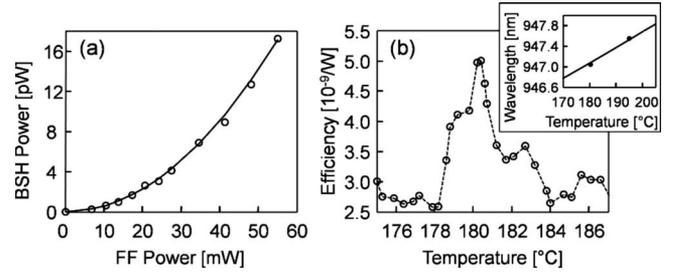

FIG. 4. (a) Measured (circles) and calculated (solid line) BSH output power vs FF excitation for the 30th resonant order. (b) Normalized conversion efficiency for the 30th BSHG order vs temperature: the input FF wavelength is 947.053 nm; the inset details a comparison between experimental data (circles) and the calculated resonant FF wavelength shift vs temperature (solid line).Appl.

The set-up for backward-SH measurements is sketched in Fig. 2. The output facet of the sample was slanted in order to decouple the forward-generated SH-back-reflected at the LN-air interface-from the sought BSHG. Nonetheless, owing to wave vector spreading, a small amount of forward-propagating SH light was generated by out-of-resonance random components of the QPM grating[19] and collected through the input lens.

Figure 3(a) shows the normalized BSHG conversion efficiency versus input FF wavelength for the 29th and 30th orders of QPM-BSHG. The different background levels at SH are ascribed to the random QPM contributions, due to domain nucleation and corresponding to spread-out periods with unequal SHG efficiencies.[20] The presence of frequency-doubled signal in both odd and even resonant orders is attributed to the unbalanced duty-cycle. Noteworthy, a duty-cycle variation of only 1.7% with respect to the ideal 50:50 can cancel out BSHG at the 29th order, letting a resonance peak emerge at the 30th [see Fig. 3(b)]. Translated in domain size, such a variation corresponds to a wall motion of ≈54 nm over the nominal coherence length of 1600 nm. The BSHG power at resonance grew quadratically with FF excitation, as graphed in Fig. 4(a).

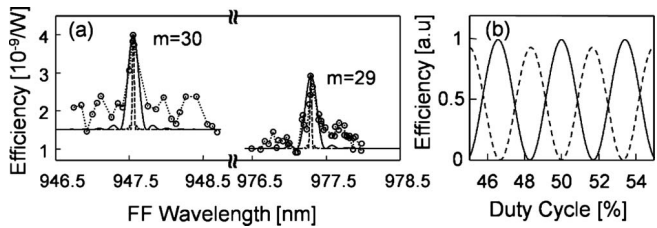

FIG. 3. (a) Normalized BSHG conversion efficiency vs FF wavelength for 29th and 30th QPM orders: data (circles) are compared with the predicted tuning for effective QPM lengths of 0.5 mm (solid lines) and 3 mm (dashed lines). (b) Calculated normalized conversion efficiency vs duty-cycle for 29th (solid line) and 30th (dashed line) at their respective resonant wavelengths.

From the BSHG data we evaluated the QPM grating in terms of period, effective length, duty-cycle, and spatial overlap between FF beam and poled domains (i.e., effective cross-section). The periodicity as extrapolated from backward SH resonance(s) is less affected by an inaccurate knowledge of the refractive index dispersion as compared to forward SHG. Moreover, when using a cw excitation the tolerances are more stringent in BSHG than in forward SHG. Therefore, BSHG measurements can be exploited as sensitive diagnostics of QPM gratings.[2,10] The spectral location of the resonances in Fig. 3(a) allowed us calculating a period of about 3.194 μm; the ratio between the two peaks yielded a duty-cycle close to 74.8:25.2. These values are consistent with those obtained from forward SHG and provide extra details in the diagnostics. From the resonant bandwidths in Fig. 3(a) we calculated an effective length of 0.5 mm (16.7% of the grating length) and, from the data in Fig. 4(a), an effective cross-section <8% of the FF beam profile. The estimates, however, are affected by (undesired) irregularities of domain walls, because the presence of random QPM lowers the overall coherence of the process, widening the resonance(s) and reducing the effective length and the conversion efficiency; hence, the values above should be considered conservative estimates as randomness was not accounted for.

Finally, Fig. 4(b) shows the normalized BSHG conversion efficiency (30th order) versus temperature, the latter permitting us a finer tuning of the resonance. The inset graphs a detail of the resonant FF wavelength shift versus temperature, the solid line being calculated from the temperature dependent Sellmeier equations for LN (as provided by the crystal supplier).

In conclusion, we reported blue-light generation in bulk PPLN via counter-propagating frequency doubling. An unbalanced duty-cycle resulted in both odd and even resonances of the quasi-phase-matching grating. A numerical assessment of the experimental data against standard SHG-BSHG model allowed us to estimate relevant parameters of the sample.

This work was supported by the Italian MIUR through PRIN 2007 Grant No. 2007CT355. GA thanks Prof. Y. J. Ding for enlightening discussions.

[1] J. A. Armstrong, N. Bloembergen, J. Ducuing, and P. S. Pershan, Phys. Rev. **127**, 1918 (1962).
[2] M. M. Fejer, G. A. Magel, D. H. Jundt, and R. L. Byer, IEEE J. Quantum Electron. **28**, 2631 (1992).
[3] S. E. Harris, Appl. Phys. Lett. **9**, 114 (1966).
[4] M. Matsumoto and K. Tanaka, IEEE J. Quantum Electron. **31**, 700 (1995).
[5] Y. J. Ding, J. U. Kang, and J. B. Khurgin, IEEE J. Quantum Electron. **34**, 966 (1998).
[6] S. Janz, C. Fernando, H. Dai, F. Chatenoud, M. Dion, and R. Normandin, Opt. Lett. **18**, 589 (1993).
[7] J. U. Kang, Y. J. Ding, W. K. Burns, and J. S. Melinger, Opt. Lett. **22**, 862 (1997).
[8] X. Gu, R. Y. Korotkov, Y. J. Ding, J. U. Kang, and J. B. Khurgin, J. Opt. Soc. Am. B **15**, 1561 (1998).
[9] X. Gu, M. Makarov, Y. J. Ding, J. B. Khurgin, and W. P. Risk, Opt. Lett. **24**, 127 (1999).
[10] X. Mu, I. B. Zotova, Y. J. Ding, and W. P. Risk, Opt. Commun. **181**, 153 (2000).
[11] C. Conti, G. Assanto, and S. Trillo, Opt. Lett. **24**, 1139 (1999).
[12] C. Conti, S. Trillo, and G. Assanto, Phys. Rev. Lett. **85**, 2502 (2000).
[13] C. Canalias, V. Pasiskevicius, M. Fokine, and F. Laurell, Appl. Phys. Lett. **86**, 181105 (2005).
[14] C. Canalias and V. Pasiskevicius, Nat. Photonics **1**, 459 (2007).
[15] L. E. Myers, Stanford University, Doctoral Dissertation (1998).
[16] A. C. Busacca, C. L. Sones, V. Apoltolopoulos, R. W. Eason, and S. Mailis, Appl. Phys. Lett. **81**, 4946 (2002).
[17] G. D. Boyd and D. A. Kleinman, J. Appl. Phys. **39**, 3597 (1968).
[18] I. Shoji, T. Kondo, and R. Ito, Opt. Quantum Electron. **34**, 797 (2002).
[19] S. Stivala, A. C. Busacca, A. Pasquazi, R. L. Oliveri, R. Morandotti, and G. Assanto, Opt. Lett. **35**, 363 (2010).
[20] A. Pasquazi, A. C. Busacca, S. Stivala, R. Morandotti, and G. Assanto, IEEE Photonics J. **2**, 18 (2010).Appl.